\documentclass[useAMS,usenatbib]{mn2e}
\usepackage{epsfig}
\usepackage{times}

\def\Msun{\mbox{M$_\odot$}}
\def\um{\mbox{$\mu$m}}
\def\ga{\mathrel{\hbox{\rlap{\hbox{\lower4pt\hbox{$\sim$}}}{\raise2pt\hbox{$>$}}}}}
\def\la{\mathrel{\hbox{\rlap{\hbox{\lower4pt\hbox{$\sim$}}}{\raise2pt\hbox{$<$}}}}}

\begin{document}

\title[The brown dwarf companion to WD\,0137-349]
{A near-infrared spectroscopic detection of the brown dwarf in the post common
  envelope binary WD\,0137-349}

\author[M. R. Burleigh et~al.]
{M. R. Burleigh$^1$ E. Hogan$^1$ P. D. Dobbie$^1$ R. Napiwotzki $^2$ 
P. F. L. Maxted$^3$\\
$^1$ Department of Physics and Astronomy, University of Leicester,
Leicester LE1 7RH, UK \\
$^2$ Centre for Astrophysics Research, STRI, University of
Hertfordshire, College Lane, Hatfield AL10 9AB, UK\\
$^3$ Astrophysics Group, Keele University, Keele, Staffordshire, ST5
5BG, UK \\
}

\date{Accepted 2100 December 32. Received 2099 December 25; 
in original form 1888 October 11}

\pagerange{\pageref{firstpage}--\pageref{lastpage}} \pubyear{2006}

\maketitle

\label{firstpage}

\begin{abstract}

We present a near-infrared spectrum of the close, detached white 
dwarf $+$ brown dwarf binary WD\,0137-349 \citep{maxted06}, 
that directly reveals the substellar
companion through an excess of flux longwards of $\approx1.95\um$. We best
match the data with a white dwarf $+$ L8 composite model. For ages 
$\sim1$~Gyr, the spectral
type of the cool secondary is in agreement with the mass determined by
\citet{maxted06} from radial velocity measurements ($0.053 \pm
0.006\Msun$), and supports an evolutionary scenario in which the brown
dwarf survived a previous phase of common envelope evolution which
resulted in the formation of this close binary. The brown dwarf is the
lowest mass companion to a white dwarf yet found, and the lowest mass
object known to have survived a common envelope phase. At $1300 < T_{\rm
  eff} < 1400$~K WD\,0137-349B is also the coolest known companion to a
white dwarf. At a separation $a = 0.65$R$_\odot$ the hemisphere
of the brown dwarf facing the $16,500$K white dwarf
is being heated through irradiation. 
We discuss the possible effects of this 
additional heating, with particular relevance to those other close
binaries with substellar companions, the hot
Jupiters. We propose future observations to investigate the likely 
temperature differences between the ``day'' and ``night'' sides of the 
brown dwarf.

\end{abstract}


\section{Introduction}

Detached brown dwarf companions to white dwarfs are rare \citep*{fbz05}. 
Proper motion surveys and searches for infra-red (IR) excesses have so far
found only three confirmed examples: GD\,165 (DA$+$L4, 
\citealt{becklin88}), 
GD\,1400 (DA$+$L$6/7$, \citealt{gd1400}; \citealt{dobbie05}), and 
WD\,0137-349 \citep{maxted06}, the subject of this paper. 
GD\,165 is a widely separated system (120\,AU); the separation
of the components in GD\,1400 is currently unknown. In contrast,
WD\,0137-349 is a close binary (orbital period $P\approx116$~minutes).  

Optical spectra of the H-rich DA white dwarf WD\,0137-349 show a narrow 
H$_\alpha$ emission line due to irradiation of the surface of the
unseen companion. Radial velocities measured from this line and the
white dwarf's intrinsic H$\alpha$ absorption line allowed 
\citet{maxted06} to determine the mass ratio of the system. Using the 
white dwarf mass ($0.39\pm0.035\Msun$), derived from an analysis of
its optical spectrum, \citet{maxted06} then determined the mass of 
the companion to be $0.053\pm0.006\Msun$. This is well below the limit
of $0.075\Msun$ commonly used to distinguish stars from brown dwarfs. 
The substellar nature of WD\,0137-349B was reinforced by an analysis of
its 2MASS near-IR $JHK$ fluxes (Figure 2, \citealt{maxted06}). 
The $J$ and $H$ band photometry can be fit by a model
consistent with the white dwarf alone. There is a slight excess of
flux at $K_{\rm S}$, which can be matched by a model consisting of the white
dwarf plus a brown dwarf spectral type mid-L or later. This is
consistent with the radial velocity determined mass measurement for
WD\,0137-349B.

Therefore, WD\,0137-349 is the first close, detached  
binary to be discovered containing a confirmed 
substellar companion. The brown dwarf must have survived a previous
phase of common envelope (CE) evolution during which it was engulfed by
the red giant progenitor of the white dwarf \citep{politano}. 
The drag on the brown
dwarf caused it to spiral in towards the red giant core from an
originally much wider orbit. Some fraction of the orbital energy was
released and deposited in the envelope, which was ejected from the
system, leaving a close binary. Simple physical arguments suggest that
low mass companions less than some limit $m_{\rm crit}$ will be
evaporated during the CE phase. WD\,0137-349B places an upper
limit on $m_{\rm crit}$. 

Alternatively, WD\,0137-349B may have originally been a planet which 
accreted a substantial fraction of its mass during the CE
phase \citep{ls83}. In this scenario, WD\,0137-349B would be
expected to have an effective temperature $T_{\rm eff}>2000$K,  
and a spectral type (early~L) 
commensurate with the cooling age of the white dwarf
($\sim0.25$~Gyr). \citet{maxted06} concluded that this model was
probably not applicable to WD\,0137-349 since the 2MASS fluxes are
inconsistent with a companion earlier than mid-L. However, 
they suggested that a definitive
test would be provided by IR spectroscopy to directly determine
an accurate spectral type for the brown dwarf.

In this paper we present a near-IR spectrum of WD\,0137-349. In
Section~2 we discuss the observations and data reduction, in Section~3
we present our analysis of these data, and in Section~4 we discuss the
results and comment on the spectral type of WD\,0137-349B.

\section{Observations and data reduction}

We were awarded Director's Discretionary Time at the 8m Gemini South 
telescope in 
November 2005, to obtain a near-IR spectrum of WD\,0137-349 with 
the Gemini Near-IR Spectrometer (GNIRS, \citealt{gnirs}) in 
programme GS-2005B-DD-9. 
We selected the cross-dispersed mode, using the
short camera, the 32~lines/mm grating centred at $1.65\um$, 
and the $0.3''$ (2
pixels) slit, which gives a resolution
$R=1700$. In this observing mode, the entire near-IR region from 
$\approx0.9\um$ to $\approx2.5\um$ is covered in a single observation with
excellent transmission across almost the whole wavelength range. 
There is no inter-order contamination as in long slit mode over a
single atmospheric window, and this mode is much more efficient than 
executing separate $H$ and $K$ band observations. 

The observations were conducted in service mode 
on the night of 2005 November 22nd. 
Our observing condition 
requirements were met: $70\%$-ile image quality (seeing 
$<0.6''$ at $J$) and $50\%$-ile cloud cover (i.e., photometric). 
The observations were obtained at a mean airmass of 1.12. 
We took $20 \times 120$s exposures, 
using two ``nod'' positions along the $6''$ long slit
giving a total exposure time of 40~minutes. This did not cause any
problems of order overlap 
during the data reduction, since the target is a point source and the 
seeing was good. With overheads for detector readout etc., the
observations lasted for $\approx55$~minutes, covering $\approx47\%$ of the
binary orbit. 
An A1\,V telluric standard was also observed (HD\,10538) both before
and after the target, at a similar airmass. 

The data were initially 
reduced using the GNIRS sub-package of the Gemini IRAF package. 
Briefly, a correction was first applied to the raw science, standard star
and arc lamp spectral images for the s-distortion in the orders. 
The data were then flat-fielded, taking care to flat-field each order
with the corresponding correctly exposed flat. 
Subsequently, difference pairs were assembled from the science and
standard star images and any significant remaining sky background
removed by subtracting linear functions, fitted in the spatial
direction, from the data. 
The spectral orders of the white dwarfs and the
standard stars were then extracted and assigned the wavelength
solution derived from the relevant arc spectrum.

Unfortunately, the GNIRS IRAF routines proved inadequate for removing
the telluric features from the target spectra and providing an 
approximate flux calibration. Instead, at this stage we utilised
standard techniques offered by software routines in the STARLINK
packages KAPPA and FIGARO. Any features intrinsic to the energy
distribution of the standard star were identified by reference to a 
near-IR spectral atlas of fundamental MK standards (e.g., 
\citealt{mkcal}; \citealt{mkcal2}; \citealt{mkcal3}) 
and were removed by linearly interpolating over them. 
The spectrum of the white dwarf was then co-aligned with the spectrum
of the standard star by cross-correlating the telluric features
present in the data. The science spectral orders were then 
divided by the corresponding standard star spectral orders and 
multiplied by a blackbody with the standard star's $T_{\rm eff}$, 
taking into account the differences in exposure times.

We found that the flux levels of the 
three brightest spectral orders (number 3, covering 
$\approx1.95\um$ to $\approx2.5\um$, number 4 covering $\approx1.4\um$ to
$\approx1.8\um$, and number 5 covering $\approx1.2\um$ to $\approx1.35\um$)
were well matched to each other. 
However, it was necessary to scale these fluxes by a single, constant 
normalisation factor 
to obtain the best possible agreement between the spectral data and
the $J$, $H$ and $K_{\rm S}$ photometric fluxes derived
from the 2MASS All Sky Data Release Point Source Catalogue magnitudes 
\citep{2MASS} where zero magnitude fluxes were taken from 
\citet{zombeck}. The reduced GNIRS spectrum and 2MASS fluxes are shown in
Figure~1. 

\section{Analysis}

\begin{figure*}
\caption{The GNIRS spectrum of WD\,0137-349 from $1.2\um - 2.4\um$ compared
to a synthetic white dwarf spectrum from a pure-H model atmosphere
normalized using the observed $V$ band magnitude (solid line). Also
shown are the synthetic white dwarf spectrum combined with spectra of
known brown dwarfs scaled to the appropriate distance as follows
(dashed lines, top-to-bottom): L0, L6, L8 and T5. Note the telluric
water bands between $1.35\um - 1.42\um$ and $1.8\um - 1.95\um$ have
been omitted.}
\label{fig:gnirs}
\psfig{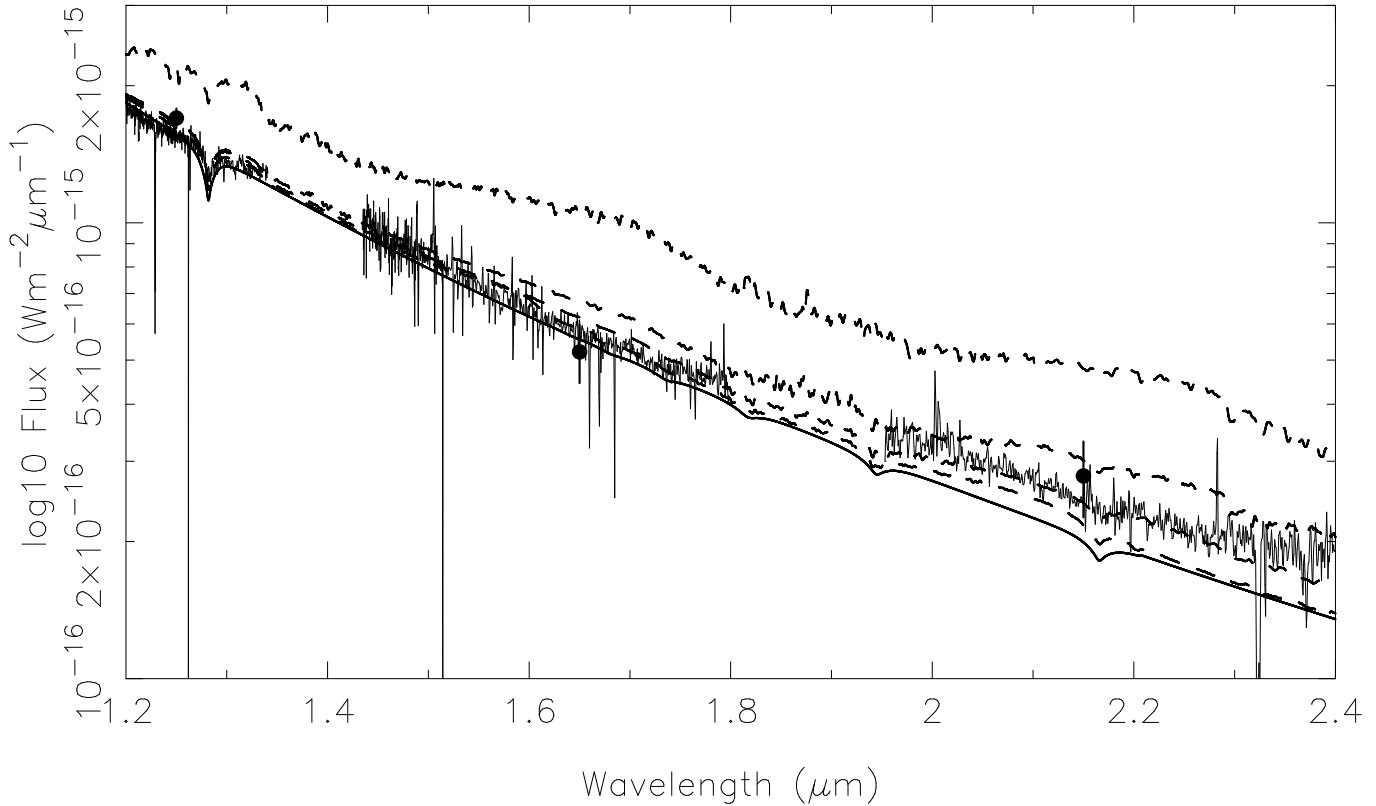}
\end{figure*}


We have analysed the GNIRS near-IR spectrum of WD\,0137-349
following the method of \citet{dobbie05}. We calculated a
pure-hydrogen 
synthetic white dwarf spectrum for the temperature ($T_{\rm eff}=16,500
\pm 500$~K) and surface gravity (log $g = 7.49 \pm 0.08$) determined by 
\citet{maxted06}. We used recent versions of the plane-parallel, 
hydrostatic, non-local thermodynamic equilibrium (non-LTE) atmosphere
and spectral synthesis codes TLUSTY (V200; \citealt{hl95} and SYNSPEC 
(v48;  
ftp:/tlusty.gsfc.nasa.gov/synsplib/synspec). The synthetic spectral
fluxes have been normalised to the white dwarf's 
$V$ magnitude ($15.33 \pm 0.02$). The synthetic white dwarf spectrum is
shown over-plotted the GNIRS data in Figure~1. 

Examination of Figure~1 reveals a clear difference in the overall
shape and level of the synthetic white dwarf spectrum and the observed
data in the $K$~band (below $\approx1.95\um$). In contrast, we do not
readily observe any spectral features typical of the energy
distributions of L or T dwarfs, e.g. Na~I absorption at 2.20$\um$,  
CH$_{4}$ or CO at 1.6$\mu$ and 2.3$\um$ respectively and 
H$_{2}$O centred on 1.15, 1.4 and 1.9$\um$. We estimate the
signal-to-noise ratio at $2.2\um\approx12$. 

To test whether the flux
excess in the $K$~band is due to the brown dwarf companion, 
and to constrain its spectral type, we have added
empirical models for substellar objects to the
white dwarf synthetic spectrum and compared these composites to the near-IR
spectrum (Figure~1). The empirical models have been constructed using the
near-IR spectra of L and T dwarfs presented by
\citet{mclean03}. In brief, the data have been obtained with the NIRSPEC
instrument on the Keck Telescope, cover the range $0.95\um -2.31\um$ with
a resolution of $\lambda/\delta\lambda\approx2000$ and have been flux
calibrated using $J$, $H$ and $K_{\rm S}$ photometric fluxes derived from
the 2MASS magnitudes as described by \citet{mclean01}. To extend
these data out to $2.4\mu$m, our effective red limit, we have appended
to them sections of UKIRT CGS4 spectra of late-type dwarfs obtained by
\citet{leggett01} and \citet{geballe02}.

The fluxes of the empirical models have been scaled to a level
appropriate to a location at d=10pc using the 2MASS $J$ magnitude of
each late-type object and the polynomial fits of 
\citet{tinney03} to the $M_{\rm J}$ versus spectral type for
L0-T8 field dwarfs/brown dwarfs. Subsequently, 
these fluxes have been re-calibrated to be consistent with the
distance of WD\,0137-349, as derived from its measured $V$ magnitude,  
effective temperature and theoretical $M_{\rm V}$ and radius. 

In Figure~1 we compare the observed GNIRS near-IR spectrum of 
WD\,0137-349 to a range of these empirical composite white dwarf $+$
brown dwarf models (WD$+$L0, L6, L8 and T5). 
We find the best match to the data is provided by 
a white dwarf $+$ L8 composite model. The lack of an obvious CO edge
at $2.3\um$ supports this conclusion.  






\section{Discussion}

We can
estimate the temperature of the brown dwarf 
from its spectral type. 
From astrometric and photometric 
observations of L~dwarfs 
\citet{vrba} give the mean effective temperature of an L8~dwarf as
  $1390$K, although they caution that their derived values should be 
treated as schematic results only, and 
note that late L and early T dwarfs appear to occupy a fairly narrow
  temperature range from $1200 < T_{\rm eff} < 1550$~K.
The temperatures computed by \citet{golimowski} for a small sample of
  late~L and early~T dwarfs are in agreement with these
  estimates. For example, they estimate the effective 
temperature of the L8 dwarf Gl\,584C, whose age is constrained by its
main sequence companion to $1 - 2.5$~Gyr \citep{Gl584C}, 
as $1300 < T_{\rm eff} < 1400$~K.  
  


Using the evolutionary models of \cite{Burrows97} and their on-line
calculator\footnote{http://zenith.as.arizona.edu/$\sim$burrows/cgi-bin/browndwarf3.cgi},
we find that for a $0.053\Msun$ 
brown dwarf to have a temperature in the range 
$1300 < T_{\rm eff} <1400$~K it must be  
$\sim1.0$~Gyr old. 
The COND models of \citet{Baraffe03} also suggest
that this mass and temperature range are consistent
only for a cooling age $\sim1.0$~Gyr.  

The low temperature of WD\,0137-349B therefore   
supports the conclusion of \citet{maxted06} that it is an
  old brown dwarf that survived a phase of CE evolution
  some $\approx0.25$~Gyr ago, and does not support the rather
  more exotic evolutionary scenario of \citet{ls83} in which a lower
  mass planet or brown dwarf accreted substantial mass during the
  CE. In that case the temperature would be substantially
  higher ($T_{\rm eff} > 2000$~K) and the spectral type would be early~L.  
 


If the brown dwarf is $\sim1.0$~Gyr old, then the main sequence
lifetime of the white dwarf progenitor was $\approx0.75$~Gyr, in which case
the progenitor's mass was just over $2\Msun$. During the CE phase,
some fraction of the orbital energy released, $\alpha_{\rm
  CE}$, will be deposited as kinetic energy in the envelope, which is
ejected from the binary.  
A progenitor mass of $2\Msun$ requires a value of $\alpha_{\rm
  CE} > 2$ to explain the formation of the WD\,0137-349 system. 
However, we note
that the cooling age of the brown dwarf does not necessarily set an
upper limit on the system age, since it may have been reheated during
the immediate post-CE phase when the white dwarf was very hot. The 
system may be older, in which case the white dwarf progenitor mass and
the required value of $\alpha_{\rm CE}$ are lower. 

The brown dwarf hemisphere facing the $16,500$K white dwarf 
is currently intercepting $\approx 1\%$ of its light. 
This irradiation is seen as 
weak ${\rm H}\alpha$~emission in the optical spectrum. 
Therefore, WD\,0137-349 is an interesting system for 
studying the effects of irradiation on the atmosphere of a 
substellar object, and can potentially 
be used as a comparison for different theoretical 
models of the effects of irradiation on lower-mass hot Jupiters. 
The heated atmospheres
of the transiting planets HD\,209458b and TrES-1 have both been detected 
by the {\it Spitzer Space Telescope} in the mid-IR during
secondary eclipse (\citealt{Deming05}, \citealt{TrES1}). Irradiation
 can increase the photospheric temperature by an order of magnitude
 compared to isolated planets, decrease the cooling rate and alter the
 planet's radius and atmospheric structure 
(e.g. \citealt{Guillot96}, \citealt{Baraffe03},
 \citealt{Arras}). Severe irradiation could 
even lead to atmospheric evaporation (\citealt{Baraffe04}), for which 
evidence has been found through the discovery of an extended
atmosphere for HD\,209458b (\citealt{Vidal03}; \citealt{Vidal04}). 
In these synchronously
rotating systems,  substantial temperature differences are expected
between the ``day'' and ``night'' sides, possibly leading to strong winds
transporting heat to the ``night'' side (\citealt{Showman},
\citealt*{Barman}). Similar effects may occur for more massive brown
dwarfs \citep*{Hubeny03}.

Neglecting the contribution from the intrinsic luminosity of the brown
dwarf, and assuming it is in thermal equilibrium with the stellar
radiation, we can use the relation given by \citet{Arras} 
to estimate the ``equilibrium'' temperature of the irradiated hemisphere:

\begin{equation}
T_{\rm eq} = T_*(R_*/2a)^{1/2}
\end{equation}
 
Taking the white dwarf radius ($R_{\rm WD}=0.0186$R$_\odot$)
and binary separation ($a=0.65$R$_\odot$) given by \citet{maxted06}, we
estimate the equilibrium temperature of the heated face 
of the brown dwarf to be $T_{\rm eq}\approx2000$K. For ages $\sim1$~Gyr, the
spectral type of an object of this temperature would be early L. 
Clearly, we are not
seeing a hemisphere at this temperature in our data. The GNIRS
spectrum was obtained across $47\%$ of the orbital period, so it is
slightly possible that the heated face has not been observed. However,
the 2MASS photometry are also not in agreement with a temperature this
high, although the phase at which they were obtained is unknown. 
Eq. (1) assumes zero reflection of the stellar photons, and
should be multiplied by $(1 - A)^{1/4}$ for an albedo $A$
(assuming isotropic emission from the brown dwarf). 
Substituting our estimated temperature
of WD\,0137-349B from the GNIRS spectrum ($T_{\rm eff}=1300$K) 
for $T_{\rm eq}$ in
Eq. (1), we find a high albedo $A\approx0.8$. For comparison,
\citet{TrES1} find that TrES-1 appears to be absorbing a high fraction
of the incident stellar flux ($A\approx0.3$), although we note that
the spectrum of the light falling on the surface of WD\,0137-349B is
very different to that falling on a hot Jupiter. 

None-the-less, there is likely some heating of the hemisphere facing the
white dwarf. A fraction of the observed flux could originate from the
heated face. This could mean that the 
underlying brown dwarf temperature might be lower
than 1300K, and the deduced age should then be regarded as a lower
limit. 

We suggest that time series K-band photometry should
reveal variability due to 
temperature differences between the ``day" and ``night" sides. 
{\it Spitzer} mid-IR observations should also reveal such
temperature variations, allowing us to characterize 
the photometric properties of the brown dwarf WD\,0137-349B across the
whole orbit. 
 

\section{Conclusions}

We have obtained a near-IR spectrum of the close, detached white 
dwarf $+$ brown dwarf binary WD\,0137-349 that reveals the substellar
companion through an excess of flux longwards of $\approx1.95\um$. We best
match the data with a white dwarf $+$ L8 composite model. This is the
lowest mass, coolest and latest spectral type companion to a white
dwarf yet discovered. For a cooling age $\sim1.0$~Gyr, the temperature of the
secondary ($1300 < T_{\rm eff} < 1400$~K) is in agreement with the
mass determined by \citet{maxted06} from radial velocity measurements
($0.053 \pm 0.006\Msun$),
and supports an evolutionary scenario in which the brown dwarf
survived a previous phase of CE evolution $\approx 0.25$~Gyr ago, 
which resulted in the formation of this close binary. The exotic
alternative evolutionary scenario of \citet{ls83}, in which a lower
mass planet or brown dwarf accreted substantial mass during the CE
phase, is not supported. If the brown dwarf is $\sim1.0$~Gyr old, then
the main sequence lifetime of the white dwarf progenitor was 
$\approx0.75$~Myr and its mass was just over 
$2\Msun$, requiring a value of the CE parameter $\alpha_{\rm CE}>2$ to
explain the formation of the system. However, the brown dwarf may be
older if it was re-heated by the very hot white dwarf immediately
after the CE phase. The brown dwarf hemisphere facing the
$16,500$K white dwarf is currently intercepting $\approx1\%$ of its
light, but does not appear to be substantially heated, although our
spectroscopic observations only cover $\approx0.5$ of the orbital period.

\section{Acknowledgments}

This paper was 
based on observations under programme GS-2005B-DD-9 
obtained at the Gemini South Observatory, which is
operated by the Association of Universities for Research in Astronomy,
Inc., under a cooperative agreement with the NSF on behalf of the
Gemini partnership: the National Science Foundation (United
States), the Particle Physics and Astronomy Research Council (PPARC, United
Kingdom), the National Research Council (Canada), CONICYT (Chile), the 
Australian Research Council (Australia), CNPq (Brazil) and CONICET
(Argentina). 

MRB \& RN acknowledge the support of PPARC Advanced Fellowships. EH 
acknowledges the support of a PPARC post-graduate studentship. PDD is
also supported by PPARC. We thank Richard Jameson for illuminating
discussions on the nature of WD\,0137-359B. We thank the staff of the 
Gemini Observatory, and Phil Lucas (University of Hertfordshire, UK)
for their advice and assistance with the GNIRS data reduction. 
We thank the referee, Kelle Cruz, for her timely and
constructive comments.

\bibliographystyle{mn2e}
\bibliography{mbu_wd0137}

\begin{thebibliography}{}

\bibitem[\protect\citeauthoryear{{Arras} \& {Bildsten}}{{Arras} \&
  {Bildsten}}{2006}]{Arras}
{Arras} P.,  {Bildsten} L.,  2006, ApJ, in press

\bibitem[\protect\citeauthoryear{{Baraffe}, {Chabrier}, {Barman}, {Allard} \&
  {Hauschildt}}{{Baraffe} et~al.}{2003}]{Baraffe03}
{Baraffe} I.,  {Chabrier} G.,  {Barman} T.~S.,  {Allard} F.,    {Hauschildt}
  P.~H.,  2003, A\&A, 402, 701

\bibitem[\protect\citeauthoryear{{Baraffe}, {Selsis}, {Chabrier}, {Barman},
  {Allard}, {Hauschildt} \& {Lammer}}{{Baraffe} et~al.}{2004}]{Baraffe04}
{Baraffe} I.,  {Selsis} F.,  {Chabrier} G.,  {Barman} T.~S.,  {Allard} F.,
  {Hauschildt} P.~H.,    {Lammer} H.,  2004, A\&A, 419, L13

\bibitem[\protect\citeauthoryear{{Barman}, {Hauschildt} \& {Allard}}{{Barman}
  et~al.}{2005}]{Barman}
{Barman} T.~S.,  {Hauschildt} P.~H.,    {Allard} F.,  2005, ApJ, 632, 1132

\bibitem[\protect\citeauthoryear{{Becklin} \& {Zuckerman}}{{Becklin} \&
  {Zuckerman}}{1988}]{becklin88}
{Becklin} E.~E.,  {Zuckerman} B.,  1988, Nature, 336, 656

\bibitem[\protect\citeauthoryear{{Burrows} et~al.,}{{Burrows}
  et~al.}{1997}]{Burrows97}
{Burrows} A.,  et~al., 1997, ApJ, 491, 856

\bibitem[\protect\citeauthoryear{{Charbonneau} et~al.,}{{Charbonneau}
  et~al.}{2005}]{TrES1}
{Charbonneau} D.,  et~al., 2005, ApJ, 626, 523

\bibitem[\protect\citeauthoryear{{Deming}, {Seager}, {Richardson} \&
  {Harrington}}{{Deming} et~al.}{2005}]{Deming05}
{Deming} D.,  {Seager} S.,  {Richardson} L.~J.,    {Harrington} J.,  2005,
  Nature, 434, 740

\bibitem[\protect\citeauthoryear{{Dobbie}, {Burleigh}, {Levan}, {Barstow},
  {Napiwotzki}, {Holberg}, {Hubeny} \& {Howell}}{{Dobbie}
  et~al.}{2005}]{dobbie05}
{Dobbie} P.~D.,  {Burleigh} M.~R.,  {Levan} A.~J.,  {Barstow} M.~A.,
  {Napiwotzki} R.,  {Holberg} J.~B.,  {Hubeny} I.,    {Howell} S.~B.,  2005,
  MNRAS, 357, 1049

\bibitem[\protect\citeauthoryear{{Elias} et~al.,}{{Elias}
  et~al.}{1998}]{gnirs}
{Elias} J.~H.,  et~al., 1998, SPIE, 3354, 555

\bibitem[\protect\citeauthoryear{{Farihi}, {Becklin} \& {Zuckerman}}{{Farihi}
  et~al.}{2005}]{fbz05}
{Farihi} J.,  {Becklin} E.~E.,    {Zuckerman} B.,  2005, ApJS, 161, 394

\bibitem[\protect\citeauthoryear{{Farihi} \& {Christopher}}{{Farihi} \&
  {Christopher}}{2004}]{gd1400}
{Farihi} J.,  {Christopher} M.,  2004, AJ, 128, 1868

\bibitem[\protect\citeauthoryear{{Geballe} et~al.,}{{Geballe}
  et~al.}{2002}]{geballe02}
{Geballe} T.~R.,  et~al., 2002, ApJ, 564, 466

\bibitem[\protect\citeauthoryear{{Golimowski} et~al.,}{{Golimowski}
  et~al.}{2004}]{golimowski}
{Golimowski} D.~A.,  et~al., 2004, AJ, 127, 3516

\bibitem[\protect\citeauthoryear{{Guillot}, {Burrows}, {Hubbard}, {Lunine} \&
  {Saumon}}{{Guillot} et~al.}{1996}]{Guillot96}
{Guillot} T.,  {Burrows} A.,  {Hubbard} W.~B.,  {Lunine} J.~I.,    {Saumon} D.,
   1996, ApJ, 459, L35

\bibitem[\protect\citeauthoryear{{Hubeny}, {Burrows} \& {Sudarsky}}{{Hubeny}
  et~al.}{2003}]{Hubeny03}
{Hubeny} I.,  {Burrows} A.,    {Sudarsky} D.,  2003, ApJ, 594, 1011

\bibitem[\protect\citeauthoryear{{Hubeny} \& {Lanz}}{{Hubeny} \&
  {Lanz}}{2001}]{hl95}
{Hubeny} I.,  {Lanz} T.,  2001, ApJ, 439, 875

\bibitem[\protect\citeauthoryear{{Kirkpatrick}, {Dahn}, {Monet}, {Reid},
  {Gizis}, {Liebert} \& {Burgasser}}{{Kirkpatrick} et~al.}{2001}]{Gl584C}
{Kirkpatrick} J.~D.,  {Dahn} C.~C.,  {Monet} D.~G.,  {Reid} I.~N.,  {Gizis}
  J.~E.,  {Liebert} J.,    {Burgasser} A.~J.,  2001, AJ, 121, 3235

\bibitem[\protect\citeauthoryear{{Leggett}, {Allard}, {Geballe}, {Hauschild} \&
  {Schwietzer}}{{Leggett} et~al.}{2001}]{leggett01}
{Leggett} S.~K.,  {Allard} F.,  {Geballe} T.~R.,  {Hauschild} P.~H.,
  {Schwietzer} A.,  2001, ApJ, 548, 908

\bibitem[\protect\citeauthoryear{{Livio} \& {Soker}}{{Livio} \&
  {Soker}}{1983}]{ls83}
{Livio} M.,  {Soker} N.,  1983, A\&A, 125, L12

\bibitem[\protect\citeauthoryear{{Maxted}, {Napiwotzki}, {Dobbie} \&
  {Burleigh}}{{Maxted} et~al.}{2006}]{maxted06}
{Maxted} P.~F.~L.,  {Napiwotzki} R.,  {Dobbie} P.~D.,    {Burleigh} M.~R.,
  2006, Nature, 422, 543

\bibitem[\protect\citeauthoryear{{McLean}, {McGovern}, {Burgasser},
  {Kirkpatrick}, {Prato} \& {Kim}}{{McLean} et~al.}{2003}]{mclean03}
{McLean} I.~S.,  {McGovern} M.~R.,  {Burgasser} A.~J.,  {Kirkpatrick} J.~D.,
  {Prato} L.,    {Kim} S.~S.,  2003, ApJ, 596, 561

\bibitem[\protect\citeauthoryear{{McLean}, {Prato}, {Kim}, {Wilcox},
  {Kirkpatrick} \& {Burgasser}}{{McLean} et~al.}{2001}]{mclean01}
{McLean} I.~S.,  {Prato} L.,  {Kim} S.~S.,  {Wilcox} M.~K.,  {Kirkpatrick}
  J.~D.,    {Burgasser} A.,  2001, ApJ, 561, L115

\bibitem[\protect\citeauthoryear{{Meyer}, {Edwards}, {Hinkle} \&
  {Strom}}{{Meyer} et~al.}{1998}]{mkcal3}
{Meyer} M.~R.,  {Edwards} S.,  {Hinkle} K.~H.,    {Strom} S.~E.,  1998, ApJ,
  508, 397

\bibitem[\protect\citeauthoryear{{Politano}}{{Politano}}{2004}]{politano}
{Politano} M.,  2004, ApJ, 604, 817

\bibitem[\protect\citeauthoryear{{Showman} \& {Guillot}}{{Showman} \&
  {Guillot}}{2002}]{Showman}
{Showman} A.~P.,  {Guillot} T.,  2002, A\&A, 385, 166

\bibitem[\protect\citeauthoryear{{Skrutskie} et~al.,}{{Skrutskie}
  et~al.}{1995}]{2MASS}
{Skrutskie} M.~F.,  et~al., 1995, in {Garzon} F.,  et~al. eds, The Impact of
  Large Scale Near-IR Sky Surveys {}.
Kluwer, Dordrecht, p.~25

\bibitem[\protect\citeauthoryear{{Tinney}, {Burgasser} \&
  {Kirkpatrick}}{{Tinney} et~al.}{2003}]{tinney03}
{Tinney} C.~G.,  {Burgasser} A.~J.,    {Kirkpatrick} J.~D.,  2003, AJ, 126, 975

\bibitem[\protect\citeauthoryear{{Vidal-Madjar} et~al.,}{{Vidal-Madjar}
  et~al.}{2004}]{Vidal04}
{Vidal-Madjar} A.,  et~al., 2004, ApJ, 604, L69

\bibitem[\protect\citeauthoryear{{Vidal-Madjar}, {Lecavelier des Etangs},
  {Désert}, {Ballester}, {Ferlet}, {Hébrard} \& {Mayor}}{{Vidal-Madjar}
  et~al.}{2003}]{Vidal03}
{Vidal-Madjar} A.,  {Lecavelier des Etangs} A.,  {Désert} J.-M.,  {Ballester}
  G.~E.,  {Ferlet} R.,  {Hébrard} G.,    {Mayor} M.,  2003, Nature, 422, 143

\bibitem[\protect\citeauthoryear{{Vrba} et~al.,}{{Vrba}  et~al.}{2004}]{vrba}
{Vrba} F.~J.,  et~al., 2004, AJ, 127, 2948

\bibitem[\protect\citeauthoryear{{Wallace} \& {Hinkle}}{{Wallace} \&
  {Hinkle}}{1997}]{mkcal}
{Wallace} L.,  {Hinkle} K.,  1997, ApJS, 111, 445

\bibitem[\protect\citeauthoryear{{Wallace}, {Meyer}, {Hinkle} \&
  {Edwards}}{{Wallace} et~al.}{2000}]{mkcal2}
{Wallace} L.,  {Meyer} M.~R.,  {Hinkle} K.,    {Edwards} S.,  2000, ApJ, 535,
  325

\bibitem[\protect\citeauthoryear{{Zombeck}}{{Zombeck}}{1990}]{zombeck}
{Zombeck} M.~V.,  1990, {Handbook of Astronomy and Astrophysics}.
Cambridge Univ. Press, UK

\end{thebibliography}

\label{lastpage}

\end{document}